%Paper: hep-th/9311170
%From: Jouko Mickesson <jouko@theophys.kth.se>
%Date: Mon, 29 Nov 93 18:32:53 EST

\input amstex.tex
\documentstyle{amsppt}
%\redefine\Bbb{\bold}
\define\a{\alpha}

\define\g{\gamma}

\redefine\l{\lambda}

\define\RM{\Bbb R}

\define\gm{\bold g}

\define\<#1,#2>{\langle #1,#2\rangle}
\define\TR{\text{tr}}
\define\dep(#1,#2){\text{det}_{#1}#2}
\magnification\magstep1
\topmatter\title RENORMALIZATION OF CURRENT ALGEBRA \endtitle
\author Jouko Mickelsson \endauthor
\affil Theoretical Physics, Royal Institute of Technology, Stockholm
S-100 44, Sweden\\ e-mail jouko\@ theophys.kth.se \endaffil
\endtopmatter
\NoBlackBoxes
\document
\baselineskip 16pt

ABSTRACT In this talk I want to explain the operator substractions needed
to renormalize gauge currents in a second quantized theory. The case of
space-time dimensions $3+1$ is considered in detail. In presence of
chiral fermions the renormalization effects a modification of the
local commutation relations of the currents by local Schwinger terms.
In $1+1$ dimensions on gets the usual central extension (Schwinger term
does not depend on background gauge field) whereas in $3+1$ dimensions
one gets an anomaly linear in the background potential.

We extend our method to the spatial components of currents. Since the
bose-fermi interaction hamiltonian is of the form $j^k A_k$ (in the
temporal gauge) we get a new renormalization scheme for the interaction.
The idea is to define a field dependent conjugation for the fermi hamiltonian
in the one-particle space such that after the conjugation the hamiltonian can
be quantized just by normal ordering prescription.
\footnote""{Presented at the conference "Generalized Symmetries in Physics"
in Clausthal, July 1993}

\vskip 0.4in
1. INTRODUCTION

\vskip 0.4in

Chiral fermions in a nonabelian external gauge field are quantized as
follows. Let $G$ be a compact gauge group, $\gm$ its Lie algebra, $M$ the
physical space, and $\Cal A$ the space of smooth $\gm$ valued vector potentials
in $M.$ For each $A\in\Cal A$ one constructs a fermionic Fock space $\Cal F_A$
containing a Dirac vacuum $\psi_A$. The Hilbert space $\Cal F_A$ carries an
irreducible representation of the canonical anticommutation relations (CAR)
$$a^*(u) a(v) + a(v) a^*(u)= (u,v) \text{ all other anticommutators $=0$}.$$
The representation is characterized by the property
$$a^*(u)\psi_A=0= a(v)\psi_A  \text{ for $u\in H_-(A)$ and $v\in H_+(A)$}
\tag1.1$$
where $H_+(A)$ is the subspace of the one-particle fermionic Hilbert space $H$
spanned by the eigenvectors of the Dirac-Weyl Hamiltonian
$$D_A =i \g_k (\nabla_k + A_k) \tag1.2$$
belonging to nonnegative eigenvalues and $H_-(A)$ is the orthogonal complement
of $H_+(A).$ Here $\nabla_k$'s are covariant derivatives in directions given
by a (local) orthonormal basis, with respect to a fixed Riemannian metric on
$M.$ In the following we shall concentrate to the physically most
interesting case dim$M=3$ and the $\g$-matrices can be chosen as the Pauli
matrices $\sigma_1,\sigma_2,\sigma_3$ with $\sigma_1 \sigma_2= i\sigma_3$
(and similarly for cyclic permutations of the indices) and $\sigma_k^2=1.$

The group $\Cal G=Map(M,G)$ of smooth gauge transformations acts on $\Cal A$
as $g\cdot A= gAg^{-1} + dg g^{-1}.$ The Fock spaces $\Cal F_A$ form a vector
bundle over $\Cal A.$ A natural question is then: How does $\Cal G$ act in
the total space $\Cal F$ of the vector bundle? Since the base base $\Cal A$ is
flat there obviously is a lift of the action on the base to the total space.
However, we have the additional physical requirement that
$$\hat g\hat D_A \hat g^{-1} = \hat D_{g\cdot A} \tag1.3$$
where $\hat D_A$ is the second quantized Hamiltonian and $\hat g$ is the lift
of $g$ to $\Cal F.$  This condition has as a consequence that $\hat g\psi_A$
should be equal, up to a phase, to the vacuum $\psi_{g\cdot A}.$

A complication in all space-time dimensions higher than $1+1$ is that the
representations of CAR in the different fibers of $\Cal F$ are inequivalent,
[A].
The effect of this is that a proper mathematical definition of the
infinitesimal
generators of $\Cal G$ (current algebra) involves further renormalizations
in addition to the normal ordering prescription. In one space dimensions the
situation is simple. The current algebra is contained in a Lie algebra
\define\gl{\bold{gl}} $\gl_1$ which by definition consists of all bounded
operators $X$ in $H$ satisfying $[\epsilon, X]\in L_2,$ where $\epsilon$ is
the sign operator $\frac{D_0}{|D_0|}$ associated to the free Dirac operator
and $L_2$ is the space of Hilbert-Schmidt operators. In general, we denote
by $L_p$ the Schatten ideal of operators $T$ with $|T|^p$ a trace-class
operator. Let $a_n^*=a^*(u_n)$, where $D_0 u_n =\l_n u_n$ and the eigenvales
are indexed such that $\l_n\geq 0$ for $n\geq 0$ and $\l_n<0$ for $n<0.$
Denoting the matrix elements of a one-particle operator $X$ by $(X_{nm})$,
the second quantized operator $\hat X$ is
$$\hat X= \sum X_{nm} :a^*_n a_m : \tag1.4$$
where the normal ordering is defined by
$$ : a^*_n a_m: =\cases -a_m a^*_n \text{ if $n=m < 0$ }\\
                 a^*_n a_m  \text{ otherwise }.\endcases$$

The commutation relations are
$$[\hat X,\hat Y]= \widehat{[X,Y]} +c(X,Y) \tag1.5$$
where $c$ is the Lundberg's cocycle, [L],
$$c(X,Y)= \frac14 \TR\epsilon [\epsilon,X][\epsilon,Y].\tag1.6$$
When $X,Y$ are infinitesimal gauge transformations on a circle
the right-hand-side
is equal to the central term of an affine Kac-Moody algebra, [PS],
$$c(X,Y)= \frac{i}{2\pi} \int_{S^1} \TR X' Y.\tag1.7$$
In this talk I want to explain the regularizations needed in $3+1$ space-time
dimensions and the generalization of (1.4) through (1.7). In section 4 we shall
use the same regularization to define a finite bose-fermi interaction
hamiltonian for QCD. (We shall not attack problems associated to vector boson
self-interactions.)

\vskip 0.4in
2. ACTION OF THE GROUP OF GAUGE TRANSFORMATIONS IN THE FOCK BUNDLE

\vskip 0.4in
Let $\epsilon(A)=\frac{D_A}{|D_A|};$ if $D_A$ has zero modes define
$\epsilon(A)$ to be $+1$ in the zero mode subspace.
For $A\in\Cal A$ denote by $P_A$ the set of unitary operators $h: H\to H$
such that
$$ [\epsilon, h^{-1} \epsilon(A) h]\in L_2.\tag2.1$$
If $h\in P_A$ then also $hs\in P_A$ for any $s\in U_1,$ where $U_1$ is the
group of unitary operators $s$ with the property $[\epsilon,s]\in L_2.$
The spaces $P_A$ form a principal bundle over $\Cal A$ with the structure
group $U_1.$

Since $\Cal A$ is flat the bundle $P$ is trivial and we may choose a section
$A\mapsto h_A\in P_A.$ Define
$$\omega(g;A)= h_{g\cdot A}^{-1} T(g) h_A\tag2.2$$
where $T(g)$ is the one-particle representation of $g\in\Cal G.$ By
construction,
$\omega$ satisfies the 1-cocycle condition
$$\omega(gg';A)= \omega(g;g'\cdot A)\omega(g';A).\tag2.3$$
Using $T(g) D_A T(g)^{-1} = D_{g\cdot A}$ we get $T(g)\epsilon(A) T(g)^{-1}=
\epsilon(g\cdot A)$ which implies
$$\align h_{g\cdot A} [\epsilon,\omega(g;A)]h_A^{-1}& = (h_{g\cdot A}\epsilon
h_{
g\cdot A}^{-1}) T(g) - T(g) (h_A\epsilon h_A^{-1})\\ & \equiv
\epsilon(g\cdot A) T(g)- T(g)\epsilon(A) \text{ mod $L_2$ }
\\& =0.\endalign $$
Since $L_2$ is an operator ideal this equation implies
$$[\epsilon, \omega(g;A)]\in L_2. \tag2.4$$
Thus the 1-cocycle $\omega$ takes values in the group $U_1.$

\bf Remark \rm In one space dimensions we can set $h_A\equiv 1$ since
$[\epsilon,
T(g)]$ is already Hilbert-Schmidt. In $d$ space dimensions the off-diagonal
blocks of $T(g)$ are only in the Schatten ideal $L_p,$ $p > d,$ [MR].

The group valued cocycle $\omega$ gives rise to a Lie algebra cocycle $\theta$
by
$$\align \theta(X;A)&= \frac{d}{dt} \omega(e^{tX};A)\vert_{t=0}\\
&= h_A^{-1} dT(X) h_A+ h_A^{-1}\Cal L_X h_A.\tag2.5\endalign $$
It satisfies the Lie algebra cocycle condition
$$\theta([X,Y];A) -[\theta(X;A),\theta(Y;A)] -\Cal L_X\theta(Y;A)+
\Cal L_Y\theta(X;A) =0,\tag2.6$$
where $\Cal L_X$ is the Lie derivative in the direction of the infinitesimal
gauge transformation $X$, $\Cal L_X f(A)= \frac{d}{dt} f(e^{-tX}\cdot A)\vert_
{t=0}.$   We denote by $dT$ the Lie algebra representation in $H$ corresponding
to the representation $T$ of finite gauge transformations. For each $A\in \Cal
A$ and $X\in Map(M,\gm)$ the operator $\theta(X;A)\in \gl_1.$

The section $h_A$ of $P$ can be used to trivialize the bundle of Fock spaces
over $\Cal A.$ Each fiber $\Cal F_A$ is identified as the free Fock space
$\Cal F_0.$ The Hamiltonian $D_A$ is quantized as
$$\hat D_A = q(h_A^{-1} D_A h_A),\tag2.7$$
that is, we first conjugate the one-particle operator $D_A$ by $h_A$ and then
canonically quantize $h_A^{-1}D_A h_A.$  The conjugated operator has a Dirac
vacuum $\psi_A$ contained in $F_0$ (but differing from the free vacuum
$\psi_0$).
The CAR algebra in the background $A$ is represented in $\Cal F_0$ through the
automorphism $a^*(u)\mapsto a^*_A(u)=a^*(h_A^{-1} u),$ $a(u)\mapsto a_A(u)=
a(h_A^{-1}
u)$ and using the free CAR representation for the operators on the right. The
Hamiltonian $\hat D_A$ is then
$$\hat D_A= \sum \l_n(A) : a_A^*(u_n) a_A(u_n):\tag2.8$$
where the $u_n$'s for nonnegative (negative) indices are the eigenvectors of
$D_A$ belonging to nonnegative (negative) eigenvalues. The normal ordering is
defined with respect to the free vacuum.

Sections of the Fock bundle are now ordinary $\Cal F_0$ valued functions.
The effect of an infinitesimal gauge transformation consists of two parts:
The Lie derivative $\Cal L_X$ acting on the argument $A$ of the function and
an operator acting in $\Cal F_0,$
$$\hat X = \Cal L_X + \sum  \theta(X;A)_{nm} : a^*_n a_m :, \tag2.9$$
where the $\theta(X;A)_{nm}$'s are matrix elements of $\theta(X;A)$ in the
eigenvector basis $(v_n)$ of $D_0.$
The commutation relations of the second quantized operators are modified by
the Lundberg's cocycle, [M1],
$$[\hat X,\hat Y]= \widehat{[X,Y]} + c(\theta(X;A),\theta(Y;A)).\tag2.10$$
In the next section we want to compute the right-hand side of (2.10) more
explicitly. We shall denote by $c_n(X,Y;A)$ ($n$=dim$M$) the second term on the
right. It is a Lie algebra 2-cocycle in the following sense:
$$c_n([X,Y],Z;A) + \Cal L_X c_n(Y,Z;A) + \text{ cyclic perm. } =0.$$

\bf Remark \rm In the case of massive Dirac fermions the cocycle vanishes
in cohomology. Namely, there is a mass gap $[-m,m]$ in the spectrum of the
Hamiltonian $D_A.$ For this reason the spaces $H_+(A)$ form a smooth vector
bundle over $\Cal A.$ Since $\Cal A$ is flat this bundle can be trivialized.
It means that one can define a continuous family of operators $h_A$ such that
$\epsilon= h^{-1}_A \epsilon(A) h_A.$ With this choice it is easy to see that
actually $[\epsilon, \omega(g;A)]=0$ and therefore the cocycle $c_n$ is
identically zero.  This does not work for massless chiral fermions because
there is no mass gap and in fact there is a spectral flow across any point
in the spectrum, that is, one can always choose a continuous path in the
space $\Cal A$ such that along the path the eigenvalues of $D_A$ crosses any
given point in the spectrum.

\vskip 0.4in
3. A COMPUTATION OF THE COCYCLE

\vskip 0.4in

Let us recall first some basic facts about pseudodifferential operators
(PSDO's). A (classical) PSDO $P$ is represented through its \it symbol. \rm
The symbol is a smooth function in the cotangent space $T^*M$ which has
an \it asymptotic expansion \rm of the form
$$p(x,\xi)= p_k(x,\xi) + p_{k-1}(x,\xi) + p_{k-2}(x,\xi)+\dots \tag3.1$$
where $n$ is an integer and the $p_j$'s are functions which are smooth
outside of the zero section in $T^*M$ and are homogeneous of degree $j$ in
the momentum variables $\xi=(\xi_1,\dots,\xi_n),$
$$p_j(x,t\xi) = t^j p_j(x,\xi) \text{ for $t> 0.$ }\tag3.2$$
The degree $k$ of the \it principal symbol \rm $p_k$ is the degree of the
PSDO $P.$  We shall consider PSDO's acting on vector valued functions. In that
case the symbols are $N\times N$ matrix valued functions. For simplicity we
shall consider only the case when the cotangent
bundle is trivial; in general, one has to cover $T^*M$ with coordinate
charts and the symbol is given by a collection of local symbols in the
coordinate charts, with appropriate rules for a change of coordinates in the
overlap sets; see [LM, III.3] for details.

A PSDO $P$ is a partial differential operator if the symbol $p$ is a polynomial
in the coordinates $\xi_j.$ In that case the operator $P$ is simply obtained
from
$p$ by replacing the coordinates $\xi_j$ by the partial derivatives $-i\partial
^x_j$ and inserting the derivatives to the right-hand-side of the coefficient
$x$-space functions.

A PSDO $P$ is defined by its asymptotic expansion up to an \it infinitely
smoothing \rm operator. An infinitely smoothing PSDO is an operator with a
symbol approaching zero faster than any power $\frac{1}{|\xi|^k}$ as $|\xi|
\to\infty.$ In particular, an infinitely smoothing operator is trace class.
A PSDO on a compact manifold of dimension $n$ is trace class if and only if
its degree $k\leq -n-1.$ The product of a pair $P,Q$ of PSDO's is represented
by the symbol
$$ (p*q)(\xi,x)= \sum_m \frac{(-i)^{|m|}}{m!} \partial_{\xi}^m p\partial_x^m q
\tag3.3$$
where the sum is over multi-indices $m=(m_1,\dots,m_n)\in \Bbb N^n,$ $|m|=
m_1 +\dots +m_n$, $m!=m_1!\dots m_n!$ and $\partial^m_x=
(\frac{\partial}{\partial x_1})^{m_1}\dots
(\frac{\partial}{\partial x_n})^{m_n}.$
In particular, the principal symbol of the product is just the (matrix) product
of the principal symbols of the factors.

In the euclidean case $M=\Bbb R^n$ a PSDO $P$ with symbol $p$ acts on sections
$\psi$ of a trivial $\Bbb C^N$
bundle over $M$ in the following way:
$$(P\psi)(x)= \int p(x,\xi) \hat{\psi} (\xi)e^{ix\cdot\xi} d^n\xi\tag3.4$$
where $\hat\psi$ is the Fourier transform of $\psi,$
$$\hat{\psi}(\xi)= \frac{1}{(2\pi)^n} \int \psi(x) e^{-i x\cdot \xi} d^n x.
\tag3.5$$
The adjoint of $P$ (in the Hilbert space of square-integrable sections, the
measure defined by a Riemannian metric on $M$) is in general a complicated
expression in terms of the symbol $p.$ We shall give the formula only in the
euclidean case:
$$P^* \sim p^* + \Omega p^* + \frac{1}{2!}\Omega^2 p^* + \dots \tag3.6$$
where
$$\Omega = -i \sum_j \partial^x_j \partial^{\xi}_j$$
and $p^*$ is the matrix adjoint of the matrix valued symbol $p.$

We shall construct the section $h_A$ explicitly as a function of the
vector potential when dim$M=3$. We shall define $h_A$ through its symbol, as a
pseudodifferential operator in the spin bundle over $M.$ I claim
that an operator with the following asymptotic expansion satisfies
the requirement (2.1):
$$h_A= 1 - \frac{i}{4}\frac{[\xi, A]}{|\xi|^2} + \text{ terms of lower order
in $|\xi|$}. \tag3.7$$
In order to make the discussion as simple as possible we
assume that $M$ is the one-point compactification of $\RM^3$ and we use
standard coordinates in $\RM^3.$  We also use the notation $A=\sum A_k
\sigma_k.$

An example of an unitary operator with the asymptotic expansion (3.1) is
the operator
$$h_A=\exp (\frac{i}{4}(D_0^2 +\l)^{-1/2} [D_0,A] (D_0^2 +\l)^{-1/2})
\tag3.8$$
where we have added a small positive constant $\l$ to the denominator
in order to cancel the infrared singularity at $\xi=0;$ this has an
effect in the asymptotic expansion only on terms of order -2 and lower
in the momentum $\xi.$ It is clear that  the lower order
terms do not have any effect on the condition (2.1) since any operator
of order $\leq -2$ is automatically Hilbert-Schmidt when the dimension of
$M$ is 3. Thus we have
$$\theta(X;A)=h_A^{-1}dT(X) h_A -h_A^{-1}\Cal L_X h_A =
X + \frac{i}{4}\frac{[\xi,dX]}{|\xi|^2} + O(-2)\tag 3.9$$
where $O(-p)$ denotes terms of order $\leq -p.$ The symbol of the PSDO
$\epsilon$ is $\frac{\xi}{|\xi|}$ and it is a simple computation to check that
indeed $[\epsilon, \theta(X;A)] \in L_2$ using the
product rule of symbols.

The term of order -2 in $\theta$ is important in computing the actual value of
the Schwinger term. It is equal to
$$\align \theta_{-2} =&-\frac14 \frac{[\sigma_k,A]}{|\xi|^2}\partial_k X
                +\frac12 \frac{[\xi,A]}{|\xi|^4}\xi_k\partial_k X\\
               & +\frac{1}{16}\frac{[\xi,A]}{|\xi|^4}
[\xi,dX].\tag3.10\endalign$$
Note that all terms are linear in the vector potential $A.$
The computation of $c_3(X,Y;A)=c(\theta(X;A),\theta(Y;A))$ is greatly
simplified when we keep in mind that it is only the cohomology class of the
cocycle $c_3$ we are interested in. Another simplification is the following:
Formally,
$$\frac14\TR\epsilon[\epsilon,P][\epsilon,Q]=-\frac12\TR[\epsilon,P]Q\tag3.11$$
when $P,Q$ are in $\gl_1.$ However, the operator on the right is not quite
trace-class; only its diagonal blocks are trace-class. For this reason
the trace is only conditionally convergent. It is convergent when evaluated
with respect to a basis compatible with the polarization $H=H_+\oplus H_-,$
for example, one can choose a basis of eigenvectors of $D_0.$
The trace of an operator $P$ with symbol $p(\xi,x)$ on a $n$-dimensional
manifold is
$$\TR P= (\frac{1}{2\pi})^n\int_{\xi,x} \TR\, p(\xi,x) d^n{\xi} d^n x
\tag3.12$$

As an exercise, let us compute (3.11) when $M=S^1$ and $P,Q$ are multiplication
operators (infinitesimal gauge transformations). In that case the symbols are
just smooth functions of the coordinate $x$ on the circle. Now $\epsilon=\frac
{\xi}{|\xi|}$ is a step function on the real line, its derivative is twice the
Dirac delta function located at $\xi=0.$  It follows that the symbol of the
commutator $\frac12[\epsilon, P]$ is
$$(-i)\delta_{\xi} p'(x) + \frac{(-i)^2}{2!} \delta'_{\xi} p''(x) +\dots.$$
Applying the formula (3.12) to (3.11) we get
$$\frac14\TR\,\epsilon[\epsilon,P][\epsilon,Q]= \frac{i}{2\pi}\int_{S^1}
\TR\, p'(x)q(x)dx, $$
where the trace under the integral sign is an ordinary matrix trace. If one
feels uneasy with singular symbols, one can approximate $\epsilon$ by a
differentiable function $\frac{\xi}{|\xi|+\l}$ and at the very end let
$\l\mapsto 0.$

In the 3-dimensional case we have to insert $P=\theta(X;A), Q=\theta(Y;A)$ in
(3.11). Using the asymptotic
expansions for $P$ and $Q,$ $p=\sum p_{-k}(\xi,x)$ one has
$$c_3(X,Y;A)= \sum_k  \TR \left(\frac{\xi}{|\xi|}*p*q-p*\frac{\xi}{|\xi|}*
q\right)_k  \tag3.13$$
In fact, one needs to take into account only finite number of terms.
The sum of terms with $k\leq -4$ is a coboundary of the 1-cochain
$$\sum_{k\geq 4} \TR\left(\epsilon* \theta(X;A)\right)_{-k}\tag3.14$$
Thus we may restrict the sum in (3.13) to indices $k>-4,$ so we have only a
finite number of
terms to check. To take care of the
infrared singularity in the integration in (3.12) we replace all denominators
$|\xi|^{-k}$ by $(|\xi|+\l)^{-k}.$ One can then check by a direct computation
that, modulo coboundaries, the result of the computation in (3.13) does not
depend on the value of $\l$ (i.e., one may take the limit $\l\mapsto 0$ in
cohomology). The final result is in accordance with the cohomological [M, F-S],
[M2], [S] and perturbative arguments, [JJ],
$$c_3(X,Y;A)= \frac{1}{24\pi^2} \int_{M} \TR A [dX,dY].\tag3.15$$

\vskip 0.4in
4. THE INTERACTION HAMILTONIAN

\vskip 0.4in

Up to this we have discussed the regularization of the time component $j_0$
(= charge density) of the nonabelian gauge current. However, in renormalized
perturbation theory one needs also the space components
$$j_k^a(x)= :\overline{\psi}(x)\gamma_k T^a\psi(x): \tag4.1$$
where the $T^a$'s are generators of $\gm.$ This is because the interaction
Hamiltonian contains the term
$$H_I= \int A^a_k(x) j^a_k(x)d^3x.\tag4.2$$
Actually, in the abelian case the hamiltonian is the free quadratic Dirac
\& Maxwell hamiltonian + the interaction $H_I.$ Thus in the abelian case
it is sufficient to renormalize $H_I$ such that it becomes a well-defined
operator in the Fock space of fermions and photons.

In this section I shall explain only the renormalizations needed to make
$H_I$ well-defined in the background quantization.

The aim is achieved through a sharpening of the regularization used for
the time component. We want to define an operator valued function $h_A$
such that
$$h_A^{-1} D_A h_A = D_0 + W_A\tag4.3$$
where $W_A$ is a PSDO of degree 0 with the additional property that
$$[\epsilon, W_A] \in L_2.\tag4.4$$
The condition (4.4) guarantees that the matrix elements
$$<\phi| W_{A^1}\dots W_{A^n}|0> \tag4.5$$
are finite, when $\phi$ is a state in the fermionic Fock space containing
a finite number of particles. Here $A^1 \dots A^n$ are any given values for
the external gauge field (smooth and with appropriate vanishing conditions
at spatial infinity when the physical space is noncompact). But the finiteness
of the matrix elements (4.5) is precisely what is needed in the perturbation
expansion, based on the Dyson expansion of the time evolution operator; see
any standard quantum field theory text book, e.g. [BD].

The choice of $h_A$ in the previous section is not quite sharp enough to
achieve
(4.4). A correct modified expression is the following:
$$\align h_A&= 1-\frac{i}{4|\xi|^2}[\xi,A]-\frac{1}{32|\xi|^4}[\xi,A]^2
-\frac18 \left[\frac{\sigma_k}{|\xi|^2}-2\frac{\xi\xi_k}{|\xi|^4},\partial_k A
\right]\\
&-\frac{1}{8|\xi|^4}[\xi,A] (A\cdot\xi) -\frac{1}{8|\xi|^4}(A\cdot \xi)[\xi,A]+
O(-3) \tag4.6 \endalign$$
After a tedious computation we obtain
$$\align W_A&=h_A^*(\xi+iA)h_A -\xi=\frac{i\xi}{|\xi|^2} A\cdot\xi \\
&-\frac18\left[\xi,[\frac{\sigma_k}{|\xi|^2} -2\frac{\xi\xi_k}{|\xi|^4},
\partial_k A]\right] -\frac{\sigma_k}{4|\xi|^2}[\xi,\partial_k A] \\
&+\frac{i}{2|\xi|^2}\epsilon_{ljk}
\xi_j[A_l,A_k]-\frac{\xi}{|\xi|^2} A_m A_m +\frac{\xi}{|\xi|^4} (A\cdot\xi)^2
+ O(-2).\tag4.7 \endalign$$
It is then a simple computation to show that $[\epsilon,W_A]$ is of degree
$-2.$
There is no magic in the derivation of the formula (4.6) for $h_A.$ It is a
simple
recursive procedure. Writing
$$h_A= 1+h_{-1} + h_{-2} +\dots$$
in the asymptotic expansion, one gets
$$h_A^* (\xi+ \a_0 +\a_{-1}+ \dots) h_A= \xi + \a_0' +\a'_{-1}+\dots,$$
where
$$\align \a'_0&= \a_0 +[\xi,h_{-1}]\\
  \a'_{-1}&= \a_{-1}+[\a_0,h_{-1}]+\xi h_{-2}+(h^*_A)_{-2}\xi
-i\sigma_k\partial
  _k h_{-1}.\tag4.8\endalign$$

The condition (4.4) is equivalent to the pair of equations
$$[\epsilon,\a'_0]=0 \text{ and } [\epsilon,\a'_{-1}]-i(\partial^{(\xi)}_k
\epsilon)( \partial^{(x)}_k\a'_0) =0$$
which together with (4.8) gives a set of linear equations for $h_{-1}$ and
$h_{-2}.$ One can then determine the lower order terms $h_k$, $k<-2$, from
the unitarity condition for $h.$ This is again a set of recursive linear
relations obtained from the formula (3.6) for the adjoint of a PSDO.

\vskip 0.4in
REFERENCES

\vskip 0.4in
[A] H. Araki in: \it Contemporary Mathematics\rm, vol. 62, American
Mathematical
Society, Providence (1987)

[BD] J. Bjorken and S. Drell: \it Relativistic Quantum Fields. \rm McGraw-Hill,
New York (1965)

[JJ] R. Jackiw and K. Johnson, Phys. Rev. \bf 182,\rm 1459 (1969)

[L] Lars-Erik Lundberg, Commun. Math. Phys. \bf 50, \rm, 103 (1976)

[LM] H. Lawson and M-L Michelsohn: \it Spin Geometry. \rm Princeton
University Press, Princeton (1989)

[M, F-S] J. Mickelsson, Commun. Math. Phys. \bf 97, \rm 361 (1985);
L. Faddeev, S. Shatasvili, Theoret. Math. Phys. \bf 60, \rm 770 (1984)

[M1] J. Mickelsson, Lett. Math. Phys. \bf 28, \rm 97 (1993)

[M2] J. Mickelsson, \it Current Algebras and Groups, \rm Plenum Press,
New York and London (1989)

[MR] J. Mickelsson and S. Rajeev, Commun. math. Phys. \bf 116, \rm 365 (1988)

[PS] A. Pressley and G. Segal, \it Loop Groups\rm, Clarendon Press, Oxford
(1986)

[S] Segal, G., preprint, Oxford 1985 (unpublished)

\enddocument